\documentclass[aps,prl,reprint,amsmath,amssymb]{revtex4-1}

\usepackage{epsfig,color,graphicx}
\usepackage{algorithm}
\usepackage{algpseudocode}

\newcommand*{\imi}{i} 
\newcommand*{\E}{\mathrm{e}}
\newcommand{\ket}[1]{\ensuremath{\vert #1 \rangle}}
\newcommand{\bra}[1]{\ensuremath{\langle #1 \vert}}
\newcommand{\braket}[2]{\ensuremath{\langle #1 \vert #2 \rangle}} 
\newcommand{\ketbra}[2]{\ensuremath{\vert #1 \rangle \langle #2 \vert}} 
\newcommand{\op}[1]{\ensuremath{\hat{#1}}} 

\begin{document}

\bibliographystyle{naturemag}

\title{
Direct unconstrained variable-metric localization of one-electron orbitals
}

\author{Ziling Luo}
\email{ziling.luo@mail.mcgill.ca}
\author{Rustam Z. Khaliullin}
\email{rustam.khaliullin@mcgill.ca}
\affiliation{Department of Chemistry, McGill University, 801 Sherbrooke St. West, Montreal, QC H3A 0B8, Canada}

\date{\today}

\begin{abstract}
Spatially localized one-electron orbitals, orthogonal and nonorthogonal, are widely used in electronic structure theory to describe chemical bonding and speed up calculations. 
In order to avoid linear dependencies of localized orbitals, the existing localization methods either constrain orbital transformations to be unitary, that is, metric preserving, or, in the case of variable-metric methods, fix the centers of nonorthogonal localized orbitals. 
Here, we developed a different approach to orbital localization, in which these constraints are replaced with a single restriction that specifies the maximum allowed deviation from the orthogonality for the final set of localized orbitals. 
This reformulation, which can be viewed as a generalization of existing localization methods, enables one to choose the desired balance between the orthogonality and locality of the orbitals. 
Furthermore, the approach is conceptually and practically simple as it obviates the necessity in unitary transformations and allows to determine optimal positions of the centers of nonorthogonal orbitals in a unconstrained and straightforward minimization procedure. 
It is demonstrated to produce well-localized orthogonal and nonorthogonal orbitals with the Berghold and Pipek-Mezey localization functions for a variety of molecules and periodic materials including large systems with non-trivial bonding. 
\end{abstract}


\maketitle

\section{Introduction} 

Spatially localized orbitals are of paramount importance in one-electron theories such as the Hartree-Fock method and Kohn-Sham density functional theory as well as in post-Hartree-Fock wavefunction-based electron correlation methods.
Localized orbitals are widely used to describe and visualize chemical bonding between atoms thus helping classify bonds and understand electronic-structure origins of observed properties of atomistic systems. 
Furthermore, localized orbitals are the key ingredient in multiple local electronic structure methods~\cite{goedecker1994efficient, bowler2012methods, zalesny2011linear, pulay1986orbital, saebo2001low, pisani2005local, hampel1996local, forner1997numerical} that dramatically reduce the computational cost of modeling electronic properties of large systems~\cite{saebo1993local, schutz1999low, hetzer2000low, schutz2001low}.
Spatially localized orbitals are known as localized molecular orbitals (LMOs) in the field of molecular quantum chemistry and maximally localized Wannier functions (MLWFs) in solid state physics and materials science~\cite{marzari2012maximally}. 
Here, they will be collectively referred to as LMOs whereas the eigenstates of the effective one-electron Hamiltonian will be called canonical molecular orbitals (CMOs) regardless of whether the system is isolated or treated with periodic boundary conditions.

In traditional localization methods, LMOs are constructed by finding a unitary transformation of CMOs that minimizes a localization function that effectively measures the spread of individual orbitals. 
Since CMOs are orthogonal and a unitary transformation preserves the orbital metric, LMOs obtained in this way are orthogonal (OLMOs) by construction~\cite{weinstein1971localized}.
Multiple localization functions have been proposed for molecular systems including Boys-Foster~\cite{boys1960construction}, Edmiston-Ruedenberg~\cite{bytautas2002electron, bytautas2003split, edmiston1963localized}, Pipek-Mezey~\cite{pipek1989a_fast}, and Von Niessen~\cite{niessen1972density}. The Boys-Foster localization~\cite{boys1960construction} is perhaps the most popular because of the simplicity of its physical interpretation, low computational complexity and ease of implementation. 
The Pipek-Mezey localization~\cite{pipek1989a_fast}, which maximizes atomic charges~\cite{mulliken1955electronic, lowdin1950non} of each orbital, is also widely used because it does not mix LMOs representing $\sigma$ and $\pi$ bonds and thus gives a clear picture of bonding patterns. 
For large supercells of condensed phase periodic systems, where electronic structure can be described with the $\Gamma$-point sampling of the Brillouin zone, several efficient optimization methods have been proposed to construct MLWF ~\cite{marzari2012maximally, resta1998quantum, resta1999electron, silvestrelli1999maximally, berghold2000general}. 
While the methods listed above rely on different localization criteria they have been formulated to ensure that orbitals remain orthogonal in a localization procedure. This is typically achieved through the exponential~\cite{berghold2000general} or Cayley parameterization of unitary transformations or through simple Jacobi rotations~\cite{edmiston1963localized}.

Due to the imposed orthogonality condition, orthogonal LMOs exhibit small non-zero values even far away from the localization center. 
These orthogonalization tails can complicate the interpretation of chemically relevant electronic-structure information and make its transferability from one system to another more difficult. 
More importantly, the tails can reduce orbital locality making orbital-based local correlation methods less computationally efficient.
To mitigate the undesirable orthogonality effects, metric-preserving unitary transformation has been applied to nonorthogonal orbitals~\cite{hoyvik2017generalising}. 
Furthermore, it has been proposed to replace unitary transformation with more general variable-metric nonsingular transformations. 
This generalization lifts the orthogonality constraint in the localization procedure and increases the number of degrees of freedom available to LMOs~\cite{anderson1968self, diner1968fully, magnasco1974localized, payne1977hartree, mehler1977self, feng2004An_efficient, cui2010efficient}. 
It has been found that NLMOs are indeed about $10-30 \%$ more localized than OLMOs if measured by the value of the Boys-Foster function~\cite{feng2004An_efficient, liu2000nonorthogonal}. 

Substantial recent efforts have been made to develop reliable algorithms to construct NLMOs~\cite{feng2004An_efficient, liu2000nonorthogonal, peng2013effective, hoyvik2017generalising}. 
Despite noticeable progress the existing methods produce NLMOs that are either still fairly similar to OLMOs~\cite{sundberg1979variationally} or lead to the linear dependence between the orbitals~\cite{feng2004An_efficient}. 
To overcome the linear dependence problem, Yang and co-workers~\cite{feng2004An_efficient, cui2010efficient} have developed a localization method in which the centers of NLMOs are fixed during the minimization of the localization function. 
The positions of the centers can be taken from the corresponding OLMOs~\cite{feng2004An_efficient} or simply guessed based on the knowledge of bonding patterns in a system~\cite{cui2010efficient}. 
While this method solves severe linear dependence issues, it requires either the computational effort to find the OLMOs centroids or good understanding of bonding properties in advance, which may limit the application of the method to relatively simple systems with unambiguous Lewis structures.

In this work, we propose a variable-metric method to simplify the construction and improve locality of LMOs. 
The key new component of the method is a simple penalty function that prevents LMOs from becoming linear dependent and allows to choose the desirable balance between the nonorthogonality and locality of LMOs. 
The penalty function replaces all constraints that are normally imposed on localized orbitals with a single restriction that specifies the allowed deviation from the orthogonality for the final localized orbitals. 
For \mbox{OLMOs}, the method is advantageous because it optimizes orbital mixing coefficients directly, obviating complicated parameterization of unitary transformations and simplifying orbital optimization algorithms.  
For \mbox{NLMOs}, the new approach allows to determine the optimal positions of their centers automatically in an unconstrained and straightforward optimization procedure, without \emph{a priori} knowledge of bonding patterns in the system.

\section{Methodology}

\subsection{Theory}

The localization procedure starts with a set of occupied one-electron states $\ket{i_0}$. 
These orbitals are not assumed to be canonical or even orthogonal. 
However, they are assumed to be normalized, which does not reduce the generality of the method. 
Furthermore, the initial orbitals must be linearly independent, that is, their overlap matrix $\sigma_{ji}^0 \equiv \braket{j_0}{i_0}$ must be invertible. 
The trial NLMOs are expressed as a linear combination of these initial states
\begin{equation}
\begin{split}
\ket{j} = \ket{i_0} {A^i}_j  
\end{split}
\end{equation}
The conventional tensor notation is used to work with the nonorthogonal orbitals~\cite{head1998tensor}: covariant quantities are denoted with subscripts, contravariant quantities with superscripts and summation is implied over the same covariant and contravariant orbital indices. Summation is not implied if two indices are both covariant. 

The loss function minimized in this work contains two terms: a conventional localization function $\Omega_L$ and a term that penalizes unphysical states with linearly dependent occupied orbitals $\Omega_P$
\begin{equation} \label{eq:fun-pen}
\begin{split}
\Omega(\mathbf{A}) = \Omega_L(\mathbf{A}) + c_P \Omega_P(\mathbf{A}), \\
\Omega_P(\mathbf{A}) = - \log \det \left[ \sigma (\mathbf{A}) \right],
\end{split}
\end{equation}
where $c_P > 0$ is the penalty strength, $\sigma$ is the NLMO overlap matrix 
\begin{equation}
\begin{split}
\sigma_{kl} = \braket{k}{l} = {A^j}_k \sigma_{ji}^0{A^i}_l
\end{split}
\end{equation}

If the NLMOs are normalized the determinant of $\sigma$ varies from 1 for orthogonal NLMOs to 0 for linearly dependent NLMOs. The penalty function---the key ingredient of the proposed method---varies from 0 to $+\infty$ for these two extreme cases, making linearly dependent state inaccessible in the localization procedure with finite penalty strength $c_P$. 
A normalization constraint can be imposed on NLMOs if their coefficients are expressed in terms of independent optimization parameters denoted with lowercase $\mathbf{a}$
\begin{equation}
\begin{split}
{A^i}_j = {a^i}_{j} ({a^k}_{j} \sigma^0_{kl}{a^l}_{j})^{-\frac{1}{2}} \equiv {a^i}_{j} N_j ,
\end{split}
\end{equation}
where the $\mathbf{a}$-dependent normalization coefficient $N_j$ is defined for brevity. 
It should be noted that the normalization requirement does not reduce the flexibility of the localization procedure and can be removed altogether if the penalty function is replaced with its more general version
\begin{equation} 
\begin{split}
\Omega_{P}^{G}(\mathbf{A}) 
 &= \Omega_{P}(\mathbf{A}) + \log \det \left[ \text{diag}(\sigma(\mathbf{A}))\right]
\end{split}
\end{equation}
with $\mathbf{A}$ now being independent optimization parameters.
%
%
It is worth mentioning that the determinant of the NLMO overlap is equal to the square of the volume of the multidimensional parallelepiped spanned by the NLMO vectors, defined by matrix $\mathbf{A}$. 

The inclusion of the penalty term converts the localization procedure into an unconstrained, straightforward optimization problem. Additionally, adjusting the strength of the penalty $c_P$ enables one to achieve the right balance between the nonorthogonality and locality of the orbitals (see below). 

\subsection{Penalty strength}

If the penalty strength $c_P$ is extremely large, $\Omega_L$ is negligible compared to the penalty term and the minimization of $\Omega$ is numerically equivalent to a trivial orbital orthogonalization. In the opposite case of extremely small $c_P$, the minimization of $\Omega$ may result in linear dependence between NLMOs reported earlier~\cite{cui2010efficient}. 

As we show below, there is a wide range of $c_P$ values between the two extremes that produce NLMOs that are substantially more localized than OLMOs and linearly independent.  
A simple strategy to find an appropriate penalty strength is to minimize $\Omega$ with a sufficiently large initial $c_P$ value and then gradually decrease $c_P$ until the determinant of the overlap of the \emph{optimal} NLMOs drops below a desired threshold $\text{D}_{\text{tar}} \in (0,1]$. 
The initial value of $c_P$ should be chosen to balance approximately the localization and penalty components of $\Omega$. 
Thus, a reasonable initial value of $c_P$ can be estimated by assuming that the reduction in $\Omega_L$ upon the minimization of $\Omega$ has the same order of magnitude as the change in the penalty $c_P \Omega_P$:
\begin{equation} \label{eq:cp-beta}
\begin{split} 
c_P^{\text{init}} & \sim \frac{ \Omega_{L}(\mathbf{I}) - \Omega_{L}(\mathbf{A}^{\ast}) }{ \Omega_{P}(\mathbf{A}^{\ast}) - \Omega_{P}(\mathbf{I}) } = \frac{ \beta_L^{\ast} }{ \beta_P^{\ast} } \times \Omega_{L}(\mathbf{I})
\end{split}
\end{equation}
where $\mathbf{A}^{\ast}$ denotes the (yet unknown) solution to the minimization problem, 
\begin{equation} 
\begin{split} 
\beta_L^{\ast} \equiv \frac{\Omega_L(\mathbf{I})- \Omega_L(\mathbf{A}^{\ast})}{\Omega_L(\mathbf{I})} \in [0,1]
\end{split}
\end{equation}
is the (positive) expected relative reduction in the localization function, and 
\begin{equation} \label{eq:betap}
\begin{split} 
\beta_P^{\ast} \equiv \log \frac{\det \sigma(\mathbf{I})}{ \det \sigma(\mathbf{A}^{\ast}) } \approx \log \frac{\det \sigma(\mathbf{I})}{ \text{D}_{\text{tar}} } \equiv \beta_P > 0
\end{split}
\end{equation}
is the logarithm of the ratio of the initial and final determinants. 
The importance of Eqs.~(\ref{eq:cp-beta})--(\ref{eq:betap}) is that they allow to estimate the initial value of $c_P$ as a product of $\Omega_L(\mathbf{I})$, which can be easily calculated in the beginning of the optimization procedure, and a dimensionless constant $\alpha$
\begin{equation} \label{eq:cp-alpha}
\begin{split}
c_{P}^{\text{init}} = \frac{ \beta_L^{\ast} }{ \beta_P } \times \Omega_L(\mathbf{I}) \equiv \alpha \times \Omega_L(\mathbf{I})
\end{split}
\end{equation}
Eq.~(\ref{eq:cp-alpha}) makes clear that the penalty component is an extensive function of a system with the units that are consistent with the localization component. Although the optimal dimensionless parameter $\beta_L^{\ast}$ is not known \emph{a priori} its magnitude can be easily estimated to obtain a sufficiently large initial guess for $c_P$. For example, an optimization of canonical orbitals $\det \sigma(\mathbf{I})=1$ that should produce the NLMO overlap determinant $\text{D}_{\text{tar}} \approx 0.1$ can be initialized by adopting the maximum possible value of $\beta_L^{\ast} = 1$, giving $\alpha = \log^{-1} 10$.

The procedure for tuning $c_P$ is shown as the outer loop of the optimization algorithm in Figure~\ref{fig:cg}. Its only required input is $\text{D}_{\text{tar}}$. It is worth mentioning that the algorithm can be modified to treat the penalty strength as a Lagrange multiplier that imposes the $\det \sigma (\mathbf{A}^{\ast}) = D_{\text{tar}}$ constraint rigorously. 

\subsection{Implementation} 

\begin{figure}
\begin{algorithm}[H]
  \caption{Conjugate gradient minimization of $\Omega$}
  \label{alg:cg}
   \begin{algorithmic}[1]
	\State Input $\epsilon_{\text{CG}}$ \Comment{Localization convergence threshold}
	\State Input $\text{D}_{\text{tar}}$ \Comment{Minimum allowed NLMO determinant}
   	\State Input $\mathbf{T}_0$ \Comment{Initial basis set coefficients for $\ket{i_0}$}
   	\State Input $\mathbf{S}$ \Comment{Basis set overlap}
   	\State Input $\mathbf{L}^K$ \Comment{Basis set representation of the localization operator} 
   	\State $\mathbf{\sigma}_0 \gets \mathbf{T}_0^{\dagger} \mathbf{ST}_0$ \Comment{Initial orbital overlap} 
   	\State $\mathbf{B}^{K} \gets \mathbf{T}_0^{\dagger} \mathbf{L}^K \mathbf{T}_0$ \Comment{Initial localization matrix, Eq.~(\ref{eq:fun-loc})} 
	\State $\mathbf{a} \gets \mathbf{I}$ \Comment{Initial guess on variational parameters}
	\State StopOuter $\gets$ False \Comment{Flag to exit the outer loop}
	\State $i_{\text{Outer}} \gets 0$ \Comment{Iteration counter}
	\Repeat \Comment{Loop to change the penalty strength}
		\State $i_{\text{Outer}} \gets i_{\text{Outer}} + 1$ 
		\State StopCG $\gets$ False \Comment{Flag to exit the CG loop}
		\State $i_{\text{CG}} \gets 0$ \Comment{Iteration counter}
		\Repeat \Comment{Fixed-penalty localization loop}
			\State $i_{\text{CG}} \gets i_{\text{CG}} + 1$ 
			\State $\mathbf{A} \gets \mathbf{a} \left[ \text{diagonal}(\mathbf{a}^{\dagger} \mathbf{\sigma}_0 \mathbf{a}) \right]^{-\frac{1}{2}}$ \Comment{Update NLMOs}
			\State $\mathbf{\sigma} \gets \mathbf{A}^{\dagger}\mathbf{\sigma}_0 \mathbf{A}$ \Comment{Update overlap}
			\State $\text{Det} \gets \text{determinant} (\mathbf{\sigma})$ \Comment{Determinant}
			\State $\Omega_{P} \gets - \log [\text{Det}] $ \Comment{Orthogonalization function}
			\State $\mathbf{P} \gets \text{Eq~(\ref{eq:grad-pen}) and~(\ref{eq:grad-convert})}$ \Comment{Orthogonalization gradient}
			\State $\Omega_{L} \gets \text{Eq~(\ref{eq:fun-loc})}$ \Comment{Localization function}
			\State $\mathbf{L} \gets \text{Eq~(\ref{eq:grad-loc}) and~(\ref{eq:grad-convert})}$ \Comment{Localization gradient}
			\If{$i_{\text{Outer}}=1$ \textbf{and} $i_{\text{CG}}=1$} 
				\State $c_{P} \gets \Omega_{L}(\log [\text{Det} / \text{D}_{\text{tar}} ])^{-1}$ \Comment{Initial strength}
			\EndIf
			\State $\Omega \gets \Omega_{L} + c_P \Omega_{P} $ 
			\If{$i_{\text{CG}}>1$}
				\State $\mathbf{\Gamma} \gets \mathbf{G}$ \Comment{Save old gradient}
			\EndIf 
			\State $\mathbf{G} \gets \mathbf{L} + c_P \mathbf{P} $ 
			\If{$\vert\vert \mathbf{G} \vert \vert_{\text{max}} < \epsilon_{\text{CG}}$}
				\State StopCG $\gets$ True
			\EndIf
			\If{\textbf{not} StopCG}
				\If{$i_{\text{CG}} > 1$}
					\State $\mathbf{O} \gets \mathbf{D}$ \Comment{Save old direction}
				\EndIf
				\State $\mathbf{D} \gets - \mathbf{G}$ \Comment{Initial direction}
				\If{$i_{\text{CG}}>1$}
					\State $\beta \gets \text{Tr}(\mathbf{G}^{\dagger} \mathbf{D})/\text{Tr}(\mathbf{\Gamma}^{\dagger}\mathbf{O})$
					\State $\mathbf{D} \gets \mathbf{D} + \beta \mathbf{O}$ \Comment{Search direction}
				\EndIf 
				\State $\alpha \gets \text{argmin}_{\alpha} \Omega(\mathbf{a} + \alpha \mathbf{D})$ \Comment{Line search}
				\State $\mathbf{a}\gets \mathbf{a} + \alpha \mathbf{D}$ \Comment{Update variational DOFs}
			\EndIf
		\Until{StopCG} 
		\If{$\text{Det} < \text{D}_{\text{tar}}$ \textbf{or} $i_{\text{Outer}} > N_{\text{Outer}}^{\text{Max}}$}
			\State StopOuter $\gets$ True
		\EndIf
		\If{$i_{\text{Outer}}>1$} 
			\State $c_{P} \gets c_P \cdot r$ \Comment{Reduce $c_P$ by factor $r \in (0,1)$}
		\EndIf
	\Until{StopOuter}
	\State $\mathbf{return}$ $\mathbf{T} \gets \mathbf{T}_0 \mathbf{A} $ \Comment{Return NLMOs coefficients}
   \end{algorithmic}
\end{algorithm}
\caption{\label{fig:cg} Algorithm for the optimization of NLMOs.}
\end{figure}

In this work, we adopted the localization function proposed by Resta~\cite{resta1998quantum, resta1999electron} and generalized by Berghold \emph{et al.}~\cite{berghold2000general} to three dimensions and simulation cells of general shape and symmetry: 
\begin{equation} \label{eq:fun-loc}
\begin{split}
\Omega_L(\mathbf{A}) &= - \sum_K \sum_i \omega_K \vert z_{i}^{K} \vert^2, \\
z_{i}^{K} &= {A^m}_i B^{K}_{mn} {A^n}_i, \\
B^{K}_{mn} &= \bra{m_0} \E^{\imi \mathbf{G}_K \cdot \mathbf{\op{r}}} \ket{n_0}
\end{split}
\end{equation}
where $\mathbf{\op{r}}$ is the position operator in three dimensions, $\omega_K$ and $\mathbf{G}_K$ are suitable sets of weights and reciprocal lattice vectors, respectively, labeled by index $K$~\cite{silvestrelli1999maximally, berghold2000general}. We chose to write the summation over $K$ explicitly because $K$ is not an orbital index. The function in Eq.~(\ref{eq:fun-loc}) can be used for both gas-phase and periodic systems~\cite{berghold2000general}. In the former case, the function is equivalent to the Boys-Foster localization~\cite{berghold2000general, resta1999electron}. In the latter case, its applicability is limited to the electronic states described within the $\Gamma$-point approximation.

We also considered the Pipek-Mezey localization function~\cite{pipek1989fast,lehtola2014pipek} that has the advantage of preserving the separation of $\sigma$ and $\pi$ bonds and is commonly employed for molecular system
\begin{equation} \label{eq:pipek}
\begin{split}
\Omega_L^{\text{PM}}(\mathbf{A}) &= - \sum_{K=1}^{\text{Atoms}} \sum_i \vert z_{i}^{K} \vert^2, \\
z_{i}^{K} &= {A^m}_i B^{K}_{mn} {A^n}_i, \\
B^{K}_{mn} &= \frac{1}{2} \sum_{\mu \in K} \bra{m_0}  \left( \ketbra{\chi_{\mu}}{\chi^{\mu}} + \ketbra{\chi^{\mu}}{\chi_{\mu}} \right) \ket{n_0}
\end{split}
\end{equation}
where $z_{i}^{K}$ is the (real) contribution of orbital $i$ to the Mulliken charge of atom $K$, $\ket{\chi_\mu}$ and $\ket{\chi^\mu}$ are atom-centered covariant and contravariant basis set functions~\cite{silvestrelli1999maximally, berghold2000general}. The summation over $\mu$ is written explicitly to emphasize that it is restricted to the basis functions centered on atom $K$. 

The unconstrained minimization of function $\Omega$ with fixed $c_P$ can be carried out with a variety of algorithms. In this work, we used a simple conjugate gradient algorithm summarized in Figure~\ref{fig:cg}. The gradient ${G_i}^j \equiv \frac{\partial \Omega}{\partial {a^i}_j}$  required in the algorithm is a sum of the localization ${L_i}^j \equiv \frac{\partial \Omega_L}{\partial {a^i}_j}$ and penalty ${P_i}^j \equiv \frac{\partial \Omega_P}{\partial {a^i}_j}$ components
\begin{equation} \label{eq:grad}
\begin{split}
G{_k}^{l} = L{_k}^{l} + c_P P{_k}^{l}.
\end{split}
\end{equation}
These components can be readily expressed in terms of the derivatives with respect to the transformation coefficients $\tilde{X}{_k}^l \equiv \frac{\partial \Omega_X}{\partial {A^k}_l}$, where $X$ is either $L$ or $P$:
\begin{equation} \label{eq:grad-convert}
\begin{split}
{X_i}^j & = \tilde{X}{_k}^l \frac{\partial {A^k}_l}{\partial {a^i}_j} = \left[ \tilde{X}{_i}^j - ( \sigma_{in}^0 {A^n}_j ) ( {A^m}_j \tilde{X}{_m}^j ) \right] N_j 
\end{split}
\end{equation}
\begin{equation} \label{eq:grad-loc}
\begin{split}
\tilde{L}{_k}^l & = - \sum_K {4 \omega_K} \times \\ 
&\times \left[  \operatorname{Re}(B^{K}_{kn}) {A^{n}}_{l} \operatorname{Re}(z_{l}^{K}) + \operatorname{Im}(B^{K}_{kn}) {A^{n}}_{l} \operatorname{Im}(z_{l}^{K}) \right]
\end{split}
\end{equation}
\begin{equation} \label{eq:grad-pen}
\begin{split}
\tilde{P}{_k}^l & = -2 \sigma_{km}^0 {A^m}_n \sigma^{nl} 
\end{split}
\end{equation}
If the gradient in Eq.~(\ref{eq:grad-loc}) is computed for the Pipek-Mezey localization function, weights $\omega_K$ are equal to one and the imaginary part of $z_{i}^{K}$ is equal to zero.

The logarithm of the determinant of the symmetric positive definite overlap matrix $\sigma$ is computed using the trace of the matrix logarithm via Mercator series
\begin{equation} \label{eq:mercator}
\begin{split}
\log \det \left( \sigma \right) &= \log \left[ \det(\mathbf{s}) \det(\mathbf{I} + \mathbf{X}) \det(\mathbf{s}) \right], \\
\log \det \left( \mathbf{I} + \mathbf{X} \right) &= \text{Tr} \log \left[ \mathbf{I} + \mathbf{X} \right], \\
\log \left[ \mathbf{I} + \mathbf{X} \right] &= \mathbf{X} - \frac{1}{2}\mathbf{X}^2  + \frac{1}{3}\mathbf{X}^3 - \frac{1}{4}\mathbf{X}^4 + \ldots
\end{split}
\end{equation}
where $\mathbf{s}=\left[\text{diag}(\sigma)\right]^{1/2}$. The series converges only if $\|\mathbf{X}\|_F < 1$. If it is not the case, the square root of $\mathbf{I}+\mathbf{X}$ is computed recursively until the norm requirement is satisfied. The advantage of this algorithm is that relies exclusively on matrix-matrix multiplication and can be readily implemented in any matrix library. Its computational cost grows cubically with the number of orbitals for dense matrices and linearly for sparse overlap matrices of localized orbitals.

The localization procedure was implemented in the CP2K software package~\cite{cp2kgeneral}. The DBCSR library~\cite{borvstnik2014sparse} that handles sparse matrices efficiently on massively-parallel computing platforms is utilized to enable orbital localization in large systems.

\subsection{Computational details}

To test the localization procedure, we constructed NLMOs for several systems ranging from a simple water molecule to complex molecules with non-trivial bonding patterns and to large periodic systems. 
For all systems, the initial CMOs were obtained using the conventional diagonalization-based SCF procedure implemented in the electronic structure module of CP2K. 
The Becke-Lee-Yang-Parr generalized gradient approximation~\cite{becke1988density, lee1988development} was used as the exchange-correlation functional.
Goedecker-Teter-Hutter pseudopotentials~\cite{goedecker1996separable} were used together with a triple-$\zeta$ atom-centered Gaussian basis set with two sets of polarization functions for all atoms. 
The energy cutoff was set at 600 Ry to define the auxiliary plane-wave basis set in the construction of the effective Hamiltonian. 
The integration over the Brillouin zone was performed using the $\Gamma$-point approximation.

\section{Results and discussion}

\subsection{Compromise between locality and orthogonality}

\begin{figure*}[hbpt]
\centering
\includegraphics[width=\textwidth]{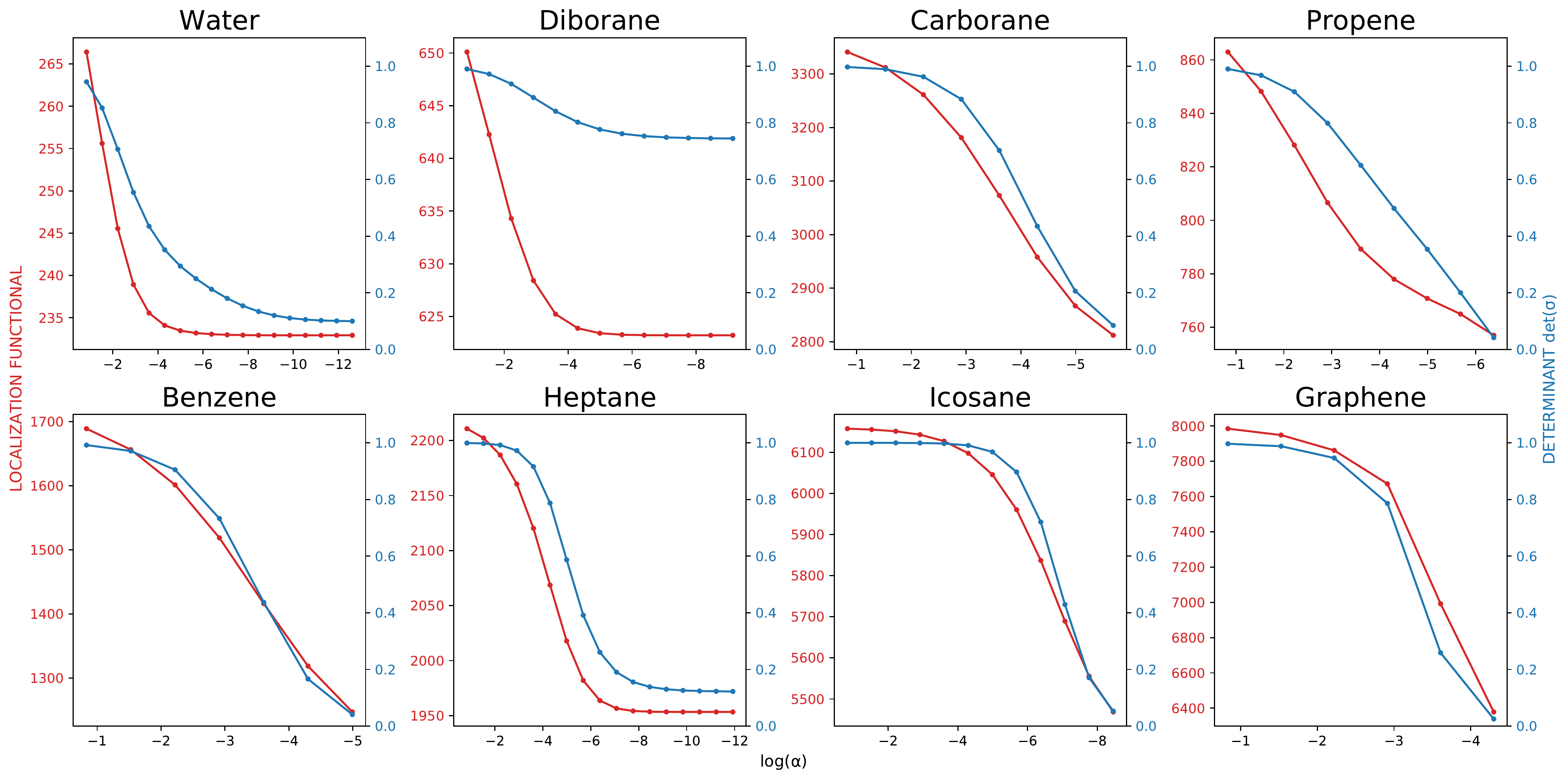}
\caption{The dependence of the optimal localization function and determinant of the NLMO overlap on $\alpha$ -- the adjustable part of the penalty strength. The first point on the left is $\alpha = \log^{-1} 10 \approx 0.434$.}
\label{fig:alpha}
\end{figure*}

For all test systems, the conjugate gradient localization procedure in Figure~\ref{alg:cg} is stable and efficient. The numerical precision of the implemented code allows to treat NLMOs with $\det(\sigma)$ as lows as $10^{-8}$ as distinct. However, a visual inspection of NLMOs with such a tiny overlap determinant reveals that many orbitals become almost identical (for example, localized on the same bonds) with only minor physically irrelevant differences. At the same time, NLMOs with $\det(\sigma) > 10^{-1}$ are found to highlight bonding patterns in all test systems correctly. Therefore, we set the minimum allowed NLMO determinant $\text{D}_{\text{tar}}$ to $10^{-1}$ in all tests. The initial value of $\alpha=\log^{-1} 10$ was set according to Eq.~(\ref{eq:cp-alpha}). The value of $\alpha$ was decreased by a factor of two in the outer loop of the algorithm (Figure~\ref{alg:cg}) until the overlap determinant fell below $\text{D}_{\text{tar}}$ or until the optimal $\Omega_{L}$ stopped changing with $c_P$ appreciably (see water, diborane, heptane examples in Figure~\ref{fig:alpha}).

Since the determinant of the overlap is proportional to the square of the volume of the parallelepiped spanned by the NLMO vectors, $\text{D}_{\text{tar}}=10^{-1}$ corresponds to the volume of $\det(\mathbf{A})\approx 0.32$, which is large enough to produce distinct NLMOs.

Figure~\ref{fig:alpha} demonstrates how the penalty strength affects the optimal orbital localization and the determinant of the orbital overlap. In all tests, the initial penalty strength is sufficiently large to produce almost perfectly orthogonal localized orbitals. At the same time it is not too large to yield more localized nonorthogonal localized orbitals after just several steps of $c_P$ adjustment. 

Figure~\ref{fig:alpha} shows that within a large range of values spanning 3--6 orders of magnitude $c_P$ serves as an adjustable parameter that can be tuned to achieve a desirable locality-orthogonality compromise. 
Thus the flexibility of the unconstrained localization method presented here allows to combine the strengths of the existing localization methods that produce either orthogonal orbitals or NLMOs with fixed localization centers. It is also important to emphasize that the localization procedure is unconstrained, does not require complicated parameterization of unitary matrices and relies on a simple easy-to-implement conjugate gradient optimization algorithm.

\subsection{NLMOs are more localized than OLMOs}

\begin{table*}[htbp]
\caption{The relative reduction in the localization function and the final determinant of the NLMO overlap.}
\label{tab:loc}
\centering
\begin{tabular}{l c c c c}
\hline\hline
Molecules & $\Delta_{\text{OLMOs/CMOs}}$  & $\Delta_{\text{NLMOs/CMOs}}$ & $\Delta_{\text{NLMOs/OLMOs}}$ & $\det(\sigma)$ \\
\hline
H$_2$O & 22 & 36 & 18 & 0.100 \\ 
CO$_2$ & 65 & 76 & 30 & 0.025 \\
Diborane (B$_2$H$_6$) & 62 & 64 & 6.2 & 0.745 \\
Borazine (B$_3$N$_3$H$_6$) & 73 & 78 & 20 & 0.026 \\
Carborane (C$_2$B$_{10}$H$_{12}$) & 72 & 76 & 17 & 0.085 \\ 
Propene (C$_3$H$_6$) & 61 & 67 & 14 & 0.042 \\
1-Butyne (C$_4$H$_6$) & 62 & 70 & 19 & 0.063 \\
Benzene (C$_6$H$_6$) & 69 & 78 & 28 & 0.041 \\ 
Heptane (C$_7$H$_{16}$) & 89 & 90 & 12 & 0.122 \\ 
Icosane (C$_{20}$H$_{42}$) & 98 & 98 & 11 & 0.053 \\ 
Decacyclene (C$_{72}$H$_{24}$) & 94 & 95 & 16 & 0.042 \\ 
Graphene & 77 & 82 & 21 & 0.025 \\
\hline
Average & 70 & 76 & 18 & 0.114 \\
\hline
\hline
\end{tabular}
\label{table:nonlin}
\end{table*}

Figure~\ref{fig:alpha} reveals the expected trend: the orbitals become more localized as they are allowed to be less orthogonal. 
The relative reduction in the localization is quantified by $\Delta_{X/Y} \equiv \frac{\Omega_L(Y)-\Omega_L(X)}{\Omega_L(Y)} \times 100 \%$, where $X$ and $Y$ can refer to CMOs, OLMOs or NLMOs.
Table~\ref{table:nonlin} compares the relative reduction as measured by the Berghold function for OLMO/CMO, NLMO/CMO and NLMO/OLMO pairs. Although NLMOs are constructed with $D_{\text{tar}}$ set to $10^{-1}$, most final values of the NLMO overlap determinants (last column in Table~\ref{tab:loc}) are somewhat lower than $D_{\text{tar}}$ because the last outer-loop iteration brings $\det(\sigma)$ below $D_{\text{tar}}$. 

The relative reduction in localization between OLMOs and CMOs ranges from 22\% to 98\% and reaches high values for extended systems (e.g. icosane) where the localization has the ability to reduce the spread significantly. The average relative reduction for the dataset considered here is 70\%. The NLMOs are even more localized than OLMOs. The relative reduction in localization between NLMOs and OLMOs ranges from 6\% to 30\% with the dataset average of 18\%. This additional reduction in the orbital spread can lead to a substantial reduction in the number of significant excitation amplitudes in local electron-correlation methods and to noticeable computational savings.

The relative reduction in localization between OLMOs and NLMOs is similar to that obtained with the algorithms that fix NLMO localization centers~\cite{feng2004An_efficient, cui2010efficient}. 
Since the NLMOs centers were previously fixed at the position of OLMOs centers this implies that the locations of the centers of NLMOs and OLMOs are very similar. It also suggests that the NLMOs obtained with $D_{\text{tar}}=10^{-1}$ represent the electronic structure of molecules reliably and also serves as an additional verification of the previously employed fixed-center procedures. 

\begin{figure}[htb]
\centering
\includegraphics[scale=0.5]{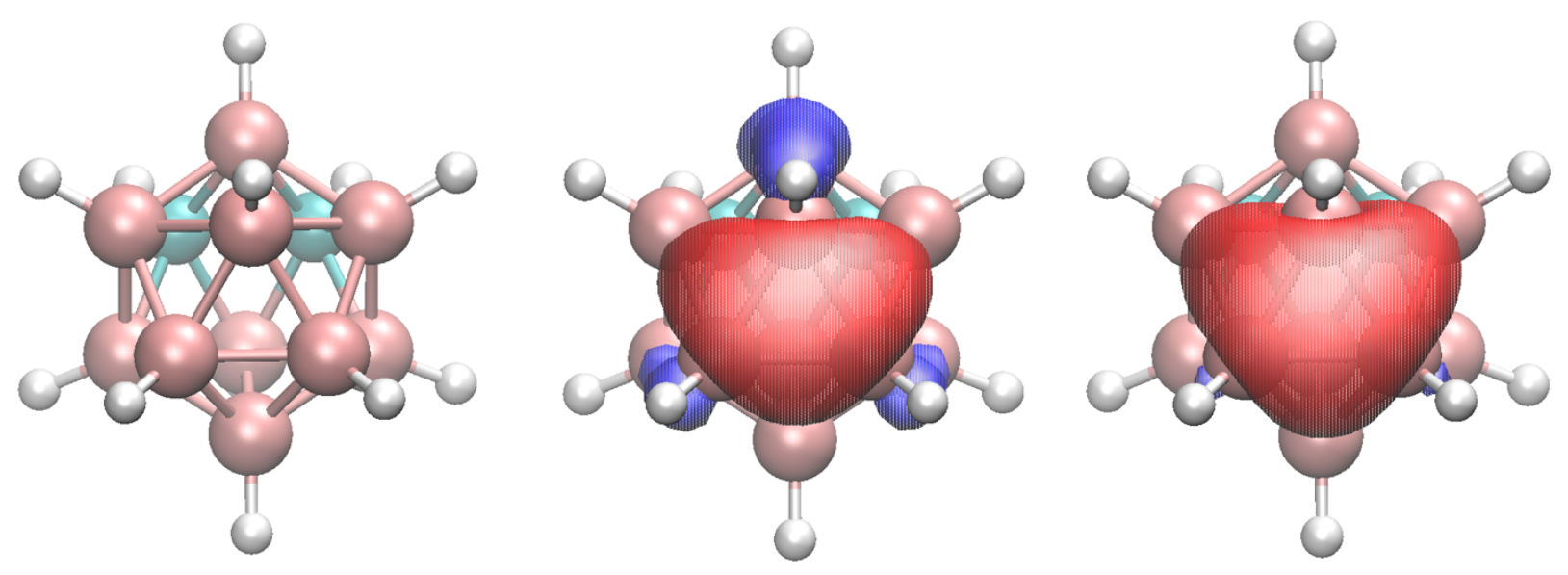}
\caption{Orthogonal (middle) and nonorthogonal (right) LMOs of three-center-two-electron B-B-B bond in the carborane molecule C$_2$B$_{10}$H$_{12}$. The isosurface value is 0.04~a.u.}
\label{fig:boro}
\end{figure}

\begin{figure}[htbp]
\includegraphics[scale=0.8]{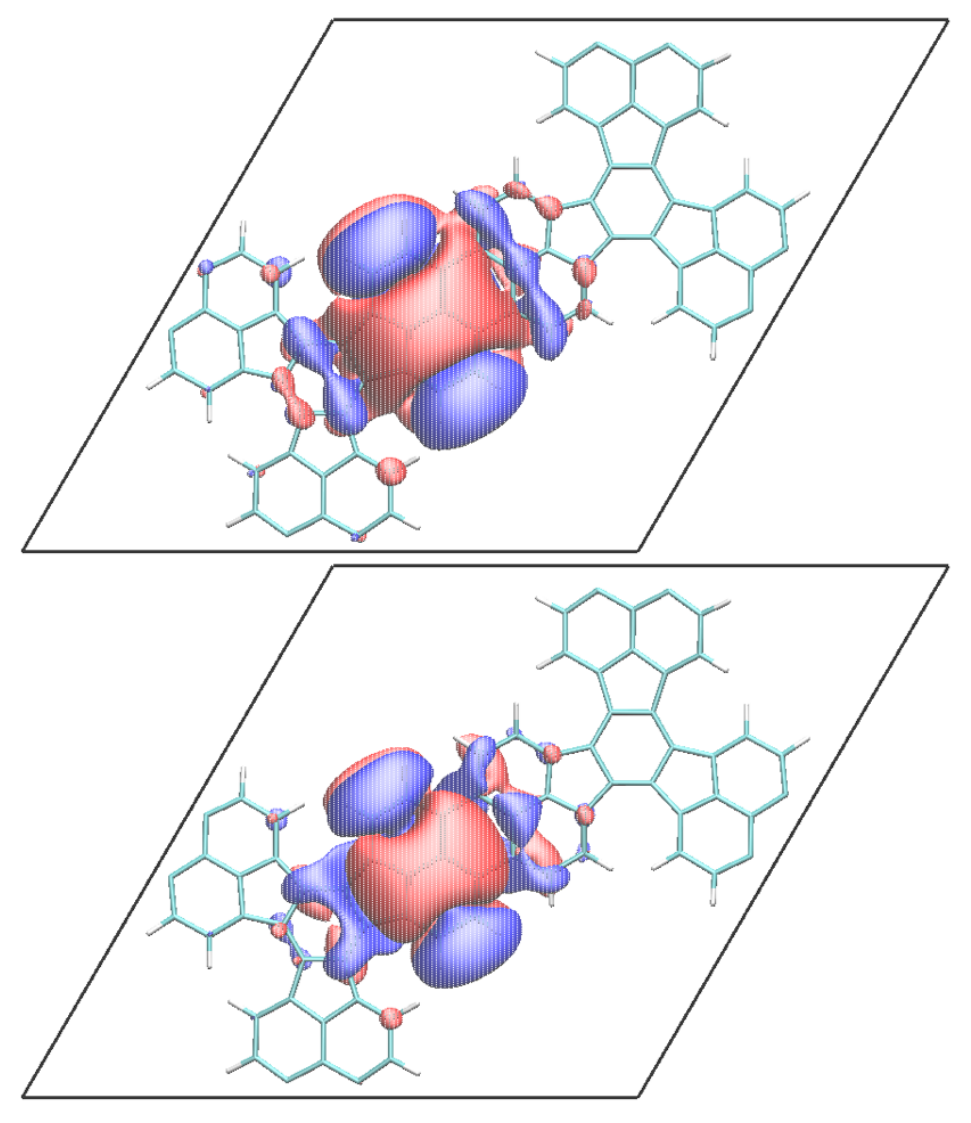} 
  \caption{OLMO (top) and NLMO (bottom) representation of a $\pi$-bond in the $\pi$-conjugated 2D polymer decacyclene C$_{72}$H$_{24}$. The isosurface value is set at a relatively low value of 0.002~a.u. to emphasize the tails of the orbitals.}
\label{fig:decacyclene}
\end{figure}

Figures~\ref{fig:boro}, \ref{fig:graphene}, \ref{fig:decacyclene} and \ref{fig:pipek} compare the shapes of NLMOs and OLMOs for several representative isolated molecules and periodic systems. 
Figure~\ref{fig:boro} shows that the NLMOs and OLMOs of a carborane C$_2$B$_{10}$H$_{12}$ molecule representing a three-center-two-electron B--B--B bond have similar positions of their centroids and similar overall shape. 
The main lobes (red region) of NLMOs are larger than those of OLMOs whereas NLMO peripheral tails are smaller. This redistribution of the probability density amplitude towards the center of NLMO is what makes NLMOs more localized than OLMOs -- the effect noted previously~\cite{liu2000nonorthogonal}. 

The reduced tails of NLMOs are also visible in Figure~\ref{fig:decacyclene}, which shows LMO representation of a $\pi$-bond in the extended $\pi$-conjugated 2D polymer decacyclene C$_{72}$H$_{24}$. Without strict orthogonality constraints, these tails can be reduced even further by imposing higher-order (e.g. quartic) penalty on amplitudes far away from the orbital center~\cite{hoyvik2012orbital}.

\subsection{$\sigma$-$\pi$ mixing}

\begin{figure}[hbpt]
\centering
\includegraphics[scale=0.6]{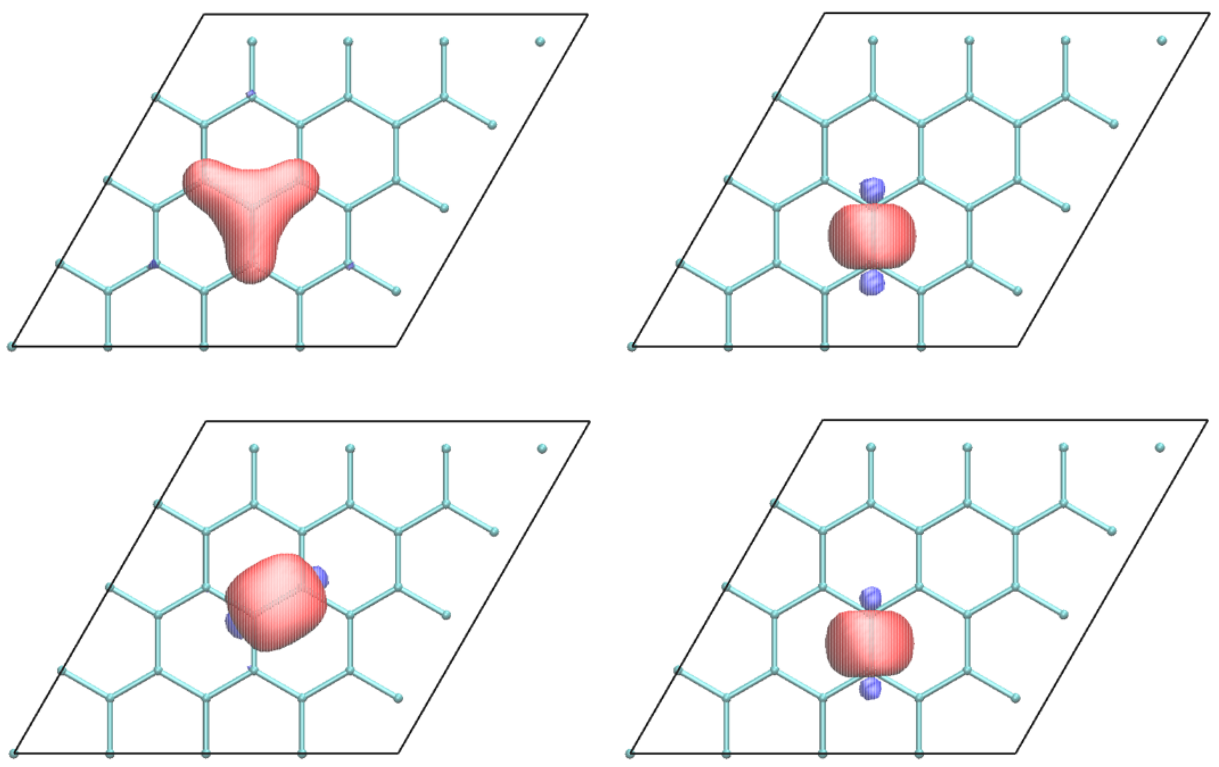}
\caption{Representative OLMOs (top) and NLMOs (bottom) of graphene. The isosurface value is 0.06~a.u.
}
\label{fig:graphene}
\end{figure}

Figure~\ref{fig:graphene} compares typical NLMOs and OLMOs of graphene. 
The single C--C $\sigma$ bonds are well reproduced by OLMOs and NLMOs but both types of LMOs fail to represent the double bonds adequately.  OLMOs tend to delocalize over several bonds complicating analysis of the electron pairs. Although NLMOs are more localized and extend only over two carbon atoms, they tend to take the shape of a mixture between $\sigma$ and $\pi$ bonds producing the so-called $\tau$ orbitals. To prevent the $\sigma$-$\pi$ mixing, the Berghold localization function was replaced with the Pipek-Mezey function. 

\begin{figure*}[htbp]
\centering
\includegraphics[scale=0.5]{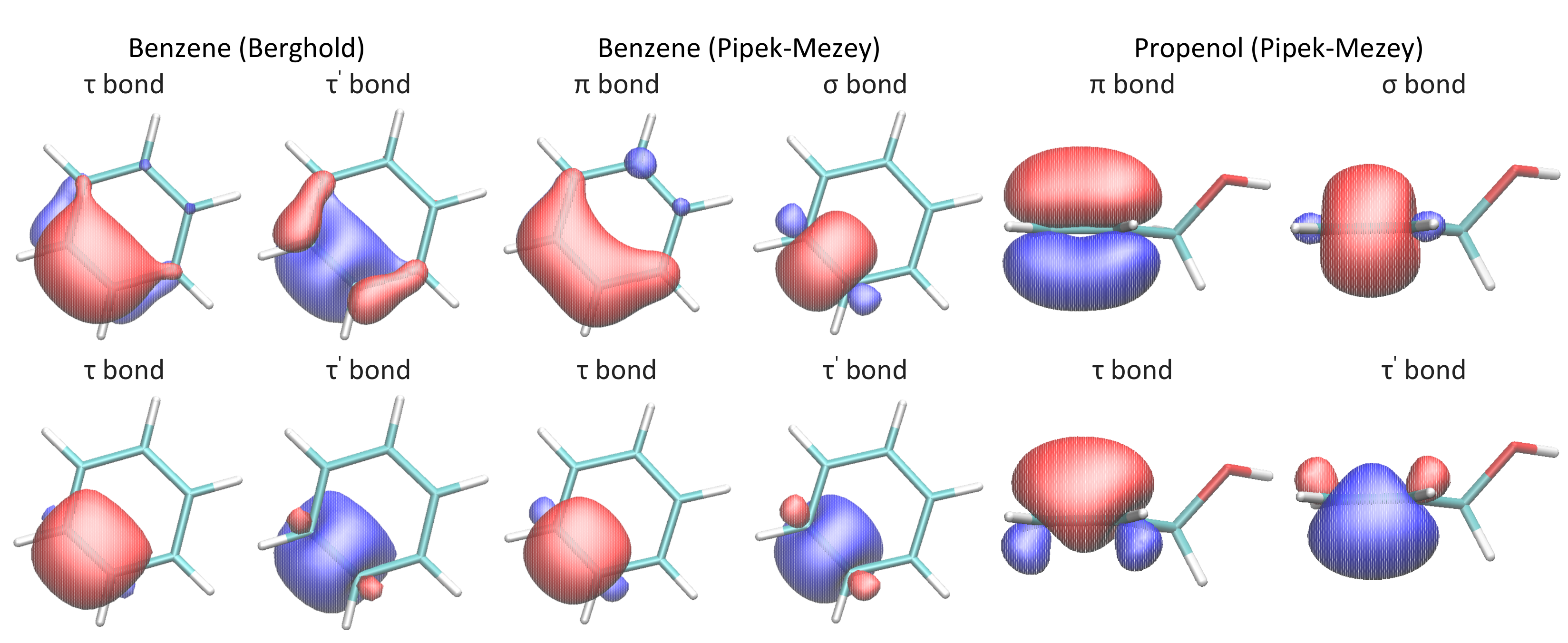}
\caption{Comparison of OLMOs (top) and NLMOs (bottom) computed with the Berghold and Pipek-Mezey localization functions for benzene and allyl alcohol. The isosurface value is 0.05~a.u.}
\label{fig:pipek}
\end{figure*}

The OLMOs and NLMOs obtained with the Pipek-Mezey and Berghold localization schemes were compared for benzene and allyl alcohol (Figure~\ref{fig:pipek}). 
For OLMOs, the Pipek-Mezey scheme preserves the separation between the $\sigma$ and $\pi$ bonds~\cite{pipek1989fast}.
However, as the determinant of the NLMO overlap decreases the $\sigma$-$\pi$ separation of the Pipek-Mezey scheme is not maintained and a pair of $\sigma$ and $\pi$ bonds tend to mix generating a pair of $\tau$ and $\tau'$ NLMOs (Figure~\ref{fig:pipek}).  Despite the failure to preserve the $\sigma$-$\pi$ separation the Pipek-Mezey NLMOs are more localized than those obtained with the Berghold (i.e. Boys-Foster) scheme.

\begin{figure}[hbpt]
\centering
\includegraphics[scale=0.35]{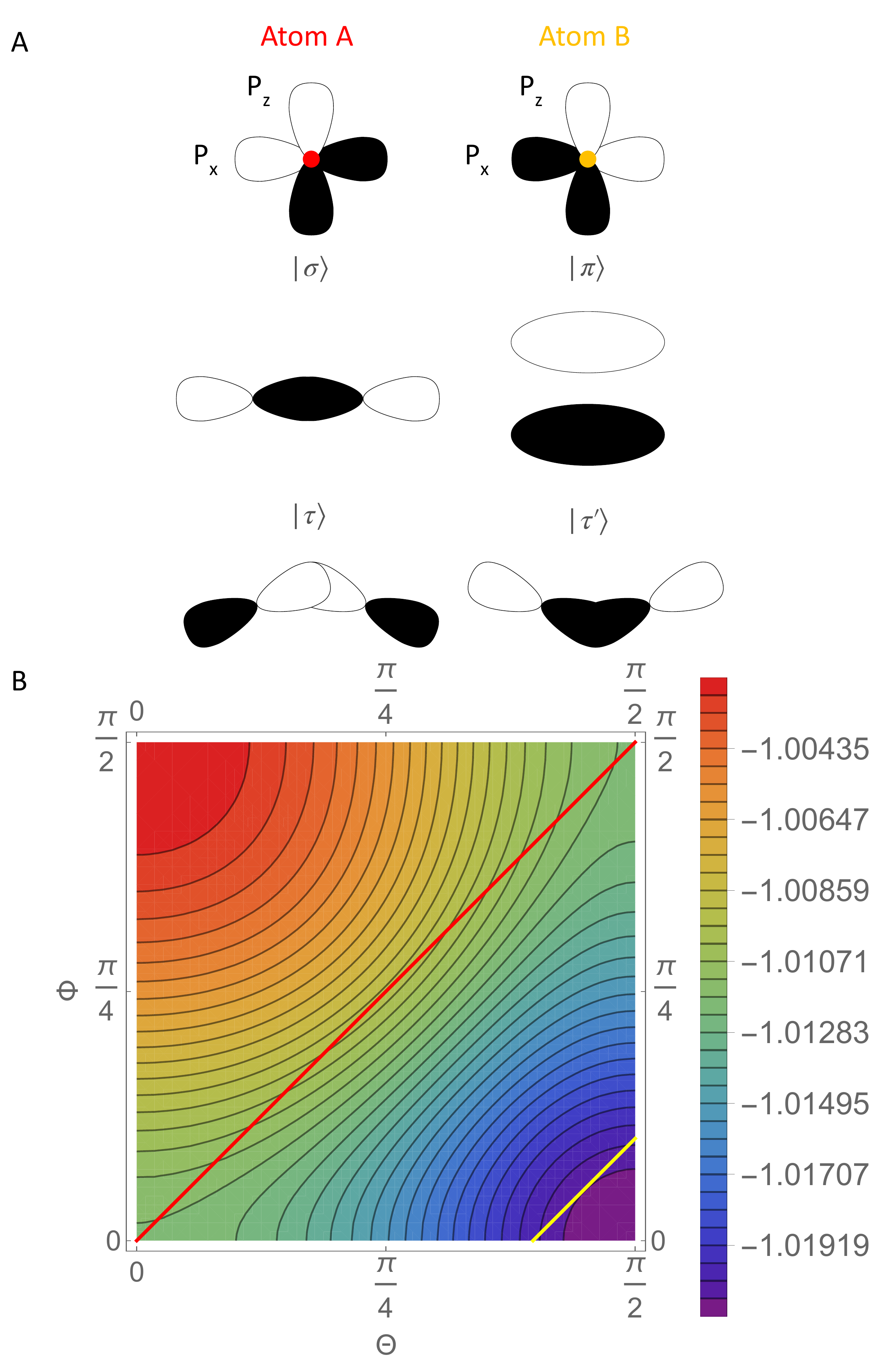}
\caption{A. The $\sigma$ and $\pi$ separation and mixed orbitals for a toy system of two atoms (A and B), each with a $p_x$ and $p_z$ atomic orbitals. B. The contour of the localization function with different value of $\Theta$ and $\Phi$. Red line: $\det(\sigma) = 1$; Yellow line: $\det(\sigma) = 10^{-1}$.}
\label{fig:sigma_pi_mixing}
\end{figure}

It has been shown that, in the original Pipek-Mezey localization scheme designed for OLMOs, $\tau$ and $\tau'$ orbitals cannot be more localized than $\sigma$ and $\pi$ orbitals~\cite{pipek1989fast}
\begin{equation} \label{eq:OLMO-pipek}
\begin{split}
\Omega_L^{\text{PM}}(\mathbf{\sigma}) + \Omega_L^{\text{PM}}(\mathbf{\pi}) \leqslant \Omega_L^{\text{PM}}(\mathbf{\tau}) + \Omega_L^{\text{PM}}(\mathbf{\tau'})
\end{split}
\end{equation}
Here we used a simple two-orbital system as an example to demonstrate that, in the case of NLMOs, the Pipek-Mezey localization function can generate $\tau$ and $\tau'$ as the most localized solution. 
We consider two atoms A and B, each with a $p_x$ and $p_z$ atomic orbitals, separated along the $x$ axis as shown in Figure~\ref{fig:sigma_pi_mixing}A.
The canonical bonding $\sigma$($\pi$)-orbital is represented by the positive linear combination of two $p_x$($p_z$) atomic orbitals. 
Mixing the occupied $\sigma$ and $\pi$ orbitals produces $\tau$ and $\tau'$ orbitals
\begin{equation} \label{eq:tao-pipek}
\begin{split}
\ket{\tau} &= \ket{\sigma}\cos(\Theta) + \ket{\pi} \sin(\Theta)\\
\ket{\tau'} &= - \ket{\sigma} \sin(\Phi) + \ket{\pi} \cos(\Phi) \\
\end{split}
\end{equation}
In the general case of NLMOs, there are two mixing angles $\Theta$ and $\Phi$ that can be varied independently to minimize $\Omega^{\text{PM}}_L$. OLMOs are recovered when $\Theta = \Phi$. 
Figure~\ref{fig:sigma_pi_mixing}B shows the dependence of the Pipek-Mezey localization function on $\Theta$ and $\Phi$. 
The red diagonal line corresponds to all possible OLMOs. The metric-preserving yellow line describes the NLMOs with $\det(\sigma) = 10^{-1}$. 
As shown in the work of Pipek and Mezey~\cite{pipek1989fast}, the $\sigma$ and $\pi$ orbitals ($\Theta=\Phi=0$) minimize the localization function along the diagonal line. 
But as the orbitals allowed to be less orthogonal, the most localized solution $\Theta+\Phi=\pi/2$ corresponds to $\tau$ and $\tau'$. 
This simple example shows that $\sigma$-$\pi$ separation should not be expected for NLMOs even if they are constructed with the Pipek-Mezey method.

\section{Conclusions}

In this work, we proposed a new approach to construct localized orthogonal and nonorthogonal one-electron orbitals. 
In this approach, the catastrophic linear dependence of the orbitals is prevented by augmenting the localization function with a simple single-value penalty that measures the degree of orbital nonorthogonality and thus allows to avoid nearly-dependent states in the localization procedure.
The proposed penalty function enables a more flexible approach to orbital localization as it allows to replace traditional metric-preserving unitary transformations of the orbitals with more general variable-metric nonsingular transformations. The approach is also conceptually simpler because a complicated parameterization of unitary transformations~\cite{hoyvik2017generalising} is obviated, even in the case of OLMOs, allowing to optimize orbital mixing coefficients directly using simple unconstrained minimization algorithms. 

The new approach is easy to implement as was demonstrated for the Berghold~\cite{berghold2000general} and Pipek-Mezey~\cite{pipek1989fast} localization functions. 
Numerous tests were performed for gas-phase and periodic systems. For gas-phase molecules, the Berghold scheme is equivalent to the Boys-Foster localization. 
For periodic systems, the electronic structure of which can be described in the $\Gamma$-point approximation, it produces orthogonal and nonorthogonal MLWFs. 

The tests show that orbital localization is fast and stable even with a simple conjugate gradient algorithm. The procedure generates NLMOs without \emph{a priori} knowledge of bonding patterns in the system. 
An additional black-box algorithm is proposed to tune the penalty strength and produce the desired balance between orthogonality and locality of NLMOs. The desired balance is specified as the minimum allowed determinant of the NLMO overlap matrix -- an intuitively clear parameter in the $(0,1]$ range. 
We found that NLMOs with the minimum allowed determinant of the order of $10^{-1}$ correctly recover bonding patterns in a variety of molecules and materials including large systems with non-trivial bonding. NLMOs are approximately $18\%$ more localized than OLMOs and have reduced tails.
This observation is consistent with previous results obtained for the NLMOs with the localization centers fixed at the OLMO positions~\cite{feng2004An_efficient, cui2010efficient}. It serves as an additional verification of the previously employed procedures and implies that the NLMOs representation of the electronic structure of molecules is reliable. 

Comparative analysis of the localized orbitals constructed with the Berghold and Pipek-Mezey localization functions indicates that the latter produces orbitals with less noticeable tails. To our surprise, the Pipek-Mezey localization, which has been a reliable method to separate orthogonal $\sigma$ and $\pi$ orbitals in locally planar systems with double bonds, tend to mix these orbitals when they are allowed to become strongly nonorthogonal. The origins of this effect was explained using a simple two-orbital system as an example.

The broader significance of this work is in the new continuous easy-to-compute nonorthogonality measure that can be adopted by a variety of electronic structure theories that benefit from nonorthogonal representation of wavefunctions. 

\section{Acknowledgments} 

The research was funded by the Natural Sciences and Engineering Research Council of Canada (NSERC) through Discovery
Grant (RGPIN-2016-0505) and by Tri-agency Institutional Programs Secretariat through New Frontiers in Research Fund (NFRFE-2018-00852). The authors are grateful to Compute Canada for computer time.

\bibliography{NLMOs}

\end{document}